# Efficient and all-carbon electrical readout of a NV based quantum sensor

Guillaume Villaret,[1,*] Ludovic Mayer,[1] Martin Schmidt,[2] Simone Magaletti,[1] Mary De Feudis,[3,] Matthew Markham,[4] Andrew Edmonds,[4] Jean-François Roch,[2] Thierry Debuisschert[1,†]

[1] Thales Research and Technology, 1 avenue Augustin Fresnel, 91767 Palaiseau Cedex, France
[2] Université Paris-Saclay, CNRS, ENS Paris-Saclay, CentraleSupelec, LuMIn, 91190 Gif-sur-Yvette, France
[3] Laboratoire de Physique des Matériaux et Surfaces, CY Cergy Paris Université, 95031 Cergy-Pontoise, France
[4] Element Six, Global Innovation Centre, Fermi Avenue, Harwell, Didcot, OX11 0QR, United Kingdom

*Abstract:*

The spin readout of an ensemble of nitrogen-vacancy (NV) centers in diamond can be realized by a photoconductive detection that is a complementary method to the optical detection of the NV electron spin magnetic resonance. Here, we implement the photoconductive detection through graphitic planar electrodes that collect the photocurrent. Graphitic electrodes are patterned using a xenon Focused-Ion Beam on an Optical-Grade quality diamond crystal containing a nitrogen concentration of ~1 ppm and a NV concentration of a few ppb. Resistance and current-voltage characteristics of the NV-doped diamond junction are investigated tuning the 532 nm pump beam intensity. The junction has an ohmic behavior and under a strong bias field, we observe velocity saturation of the optically-induced carriers in the diamond junction. We perform the photoconductive detection in continuous-wave regime of the magnetic resonance of the NV centers ensemble for a magnetic field applied along the <100> and the <111> direction with a magnitude above 100 mT. This technique enables the realization of all-carbon diamond quantum sensors integrating graphitic microstructures for the electrical readout.

Nitrogen-vacancy (NV) center is a spin defect in diamond whose appealing properties make it a promising candidate for quantum sensing[1]. The standard detection scheme used for monitoring the magnetic resonance of the NV center electron spin relies on an optical readout and is therefore constrained by the high refractive index of diamond, which drastically affects the photon collection efficiency and consequently the sensitivity of all optical NV-based sensing techniques[2]. Alternatively, the NV magnetic resonance can be detected by the photocurrent produced by an ensemble of NV centers[3,4,5,6,7,8], down to the single defect level[9]. This electrical detection can lead to compact integrated NV-based sensors. Contrary to the optical readout, the collection of charge carriers produced by the NV spin resonance is not limited by the saturation related to the optical transition[9], thus possibly leading to an improvement of the magnetic field sensitivity. However, engineering efficient electric contacts for the charge collection at the surface of the diamond sample is challenging. Fabrication of ohmic contacts on diamond usually relies on forming covalent carbide bonding between a stacked metallic layer and the diamond substrate using metallic layer deposition followed by annealing[10,11,12]. The electrodes can also be realized by graphitizing the diamond itself using thermal annealing[13], laser irradiation[14,15] or high-energy ion implantation[16,17] on the diamond surface or in the bulk[18]. Graphitization of diamond makes it possible to produce electrodes with both ohmic contact and lower contact resistance than with metallic deposition[19]. Furthermore, graphitization alleviates the adhesive contact problem on the inert diamond surface[14], which strongly influences the good performance of the device and its lifetime. To the best of our knowledge, only the first approach based on metallic deposition has been implemented for the photoconductive detection of the magnetic resonance in experiments based on ensembles of NV centers. Here, we report the electrical readout of the spin resonance of an ensemble of NV centers using graphitic electrodes that collect the optically induced charge carriers.

[*] Present address: Pasqal, 7 Rue Léonard de Vinci 91300 Massy, France
[†] Author to whom correspondence should be addressed: thierry.debuisschert@thalesgroup.com

The NV center in diamond consists of the association of a substitutional nitrogen impurity and a nearby carbon vacancy in the diamond lattice. According to the tetrahedral structure of the diamond lattice, four different N-to-V axis orientations are possible. Depending on the number of electrons trapped in the point defect, the NV center has two charge states with specific photophysical and spin properties (Figure 1). In the neutral charge state $NV^0$, the NV center contains five electrons leading to ground and excited spin doublet states noted $^2E$ and $^2A_1$.[20] Under optical excitation, the $NV^0$ charge state emits an intense photoluminescence (PL) signal with a zero-phonon line (ZPL) at 2.15 eV (575 nm) and a broad phonon sideband.[21] The capture of an extra electron leads to the negative charge state $NV^-$ with an energy level structure of two spin-triplet states (ground state $^3A_2$ and excited state $^3E$) and two spin-singlet states (lower state $^1E$ and a higher-lying state $^1A_1$). Under optical excitation, spin-preserving (i.e. $\Delta m_s = 0$ where $m_s = 0, \pm 1$ is the electron spin projection along the N-V intrinsic quantization axis of the defect) optical transitions occur between $^3A_2$ and $^3E$ states. These transitions lead to a PL signal slightly red-shifted compared to the $NV^0$ PL with a ZPL at 1.945 eV (637 nm) and a broad phonon sideband.[21,22] Moreover, a spin-selective and nonradiative decay path connects the excited state $^3E$ to the higher singlet state $^1A_1$ and the lower singlet state $^1E$ to the ground triplet state $^3A_2$.[23,24] This intersystem crossing induces the optical polarization of the spin into the brighter sublevel $m_s = 0$ of the $NV^-$ ground state $^3A_2$, and makes possible the optical readout of the $NV^-$ spin state exploiting the lower PL rate of the $NV^-$ under illumination when the sublevels $m_s = \pm 1$ are populated by the application of a resonant microwave (MW) magnetic field excitation.[25] These properties allow performing the optical detection of the magnetic resonance (ODMR) between the $m_s = 0$ and the $m_s = \pm 1$ ground states.[26]

The alternative photoelectric detection of the magnetic resonance (PDMR) of the $NV^-$ center is based on the electric signal resulting from the photoionization of the $NV^-$ charge state into the conduction band (CB) of the crystal.[27,28,29] Under optical excitation at a wavelength of 532 nm (2.33 eV), the photoionization of the NV⁻ center occurs mainly through the following two-photon process (Figure 1). A first photon induces the optical transition from the ground state $^3A_2$ to the excited state $^3E$, then a second photon promotes an electron from the excited state $^3E$ to the conduction band located 0.7 eV above, turning the NV⁻ into the NV⁰ charge state.[30] For the same reason that the spin-selective intersystem crossing at the excited state $^3E$ creates a spin-dependent PL intensity, the photoionization rate of the NV⁻ from $^3E$ excited state, denoted $\Gamma_e$ on Figure 1, has a similar spin dependence (i.e. $\Gamma_e$ is higher from the sublevel $m_s = 0$ than from the sublevels $m_s = \pm 1$):

$$NV^{-*} \xrightarrow{\Gamma_{e,spin}} NV^0 + e^- \qquad (1)$$

The back-conversion process of NV⁰ to NV⁻ is also a two-photon process that leads to a steady state between the two charge states. A first photon induces a transition from the ground state $^2E$ to the excited state $^2A$ of the NV⁰ charge state, then a second photon promotes an electron from the valence band (VB) located 1.21eV below to the empty $^2E$ level. This reverse process from NV⁰ to NV⁻ leaves a free hole in VB with a rate $\Gamma_h$.[31]

Using this stable conversion scheme, the NV⁻ spin resonance can be electrically detected by monitoring the photocurrent induced between two electrodes (Figure 2a). The PDMR signal is therefore an alternative to the ODMR detection scheme.[3]

The diamond is a single-crystal diamond sample of lateral dimension 3.2 x 3.2 mm² and a thickness of 0.5 mm. The substrate was Electronic Grade high purity diamond produced by means of chemical vapor deposition (CVD) process. The substrate was grown on along a <100> direction with a N doping concentration of around 1 ppm. The crystal was cut vertically out of sample, processed and optically polished leading to an Optical Grade quality diamond crystal with a {110} major face, <110> and <100> edges and a [NV⁻] concentration of a few ppb measured via confocal microscopy using a NV⁻ specific filter (FF01-697/75 nm band-pass filter, Semrock).

Graphitic electrodes were patterned on the {110} top face of the diamond sample using Focused-Ion Beam (FIB) with a plasma source filled by xenon gas.[32] A dose of $2 \cdot 10^{15}$ singly charged xenon atoms (Xe⁺) per cm² with 30 keV kinetic energy was implanted at a depth of about 10 nm below the diamond surface (Figure 2b, lower). The resulting defects destroy the diamond lattice, which relaxes into an amorphous graphitic structure for a vacancy density above the damage threshold of $10^{22}$ vacancies (Vac) per cm³.[33] Since the ion energy, the dose and the damage vary with distance from the surface, this irradiation creates a gradual transition from graphite to diamond and provides good electrical contact with high mechanical adhesion. Exploiting the spatial resolution of the FIB irradiation, four different sets of electrode geometry were realized. Each electrode has a rectangular shape of 50 x 100 μm² while the gap between two electrodes was 5, 7.5, 10, and 20 μm, respectively (Figure 2b, upper). We then applied an external voltage $U$ to the electrodes that generates a static electric field $E$ to collect both electrons ($e^-$) and holes ($h^+$) from ionized and back-converted NV centers under optical excitation.

Considering a homogeneous charge carrier density and electric field in the medium between the electrodes, the circulating steady-state photocurrent density $j$ is:

$$j = e \cdot (nv_e + pv_h) = e \cdot (\tau_e \Gamma_e \mu_e + \tau_h \Gamma_h \mu_h) \cdot E = \sigma \cdot E \qquad (2)$$

where $e$ is the elementary charge (C), $n$ is the electron carrier concentration (m$^{-3}$), $p$ is the hole carrier concentration (m$^{-3}$), $v_e$ and $v_h$ are the charge carriers drift velocities for respectively electrons and holes (m.s$^{-1}$), $\tau_e$ and $\tau_h$ are the charge carriers lifetimes (s), $\Gamma_e$ and $\Gamma_h$ are the charge carriers generation rates per unit of volume (s$^{-1}$.m$^{-3}$), $\mu_e$ and $\mu_h$ are the charge carrier drift mobilities (m$^2$.V$^{-1}$.s$^{-1}$), $E$ is the electric field (V.m$^{-1}$) and $\sigma$ denotes the conductivity (A.V$^{-1}$.m$^{-1}$) of the NV-activated diamond sample. Therefore, the change in the ionization rate of the NV center induced by the MW excited spin resonance can be detected by monitoring the change in the photoconductivity between the graphitic electrodes.[3]

We first characterize the electrical behavior of each graphitic electrode by a two-point probe method using a Keithley 2420 source measure unit and two tungsten probes for contact. We measure a resistance $R_C$ ~240 kΩ for an electrode length of 100 μm (Figure 2c). This high resistance of the electrodes is mainly due to the contact resistance between the tips and the graphitic layer since it does not depend on the distance between the tips. We then use the graphitic electrodes to perform a photoconductive measurement of the NV-doped diamond junction under green illumination (Figure 2d). The NV centers are optically excited using a continuous wave (CW) pump laser at 532 nm wavelength (Cobolt Samba 1 W). The laser beam is focused on the side of the diamond crystal through a {110} facet with a long working distance objective (Mitutoyo, M Plan Apo 20X, 0.42 N.A.) giving a spot size of 1.5 μm a few microns below the diamond surface. By varying the applied voltage between the electrodes, the photocurrent in the junction goes from an ohmic regime for voltages below about 10 V to a saturation regime reached for higher voltages (Figure 2d). Even for high optical excitation power, the resistance of the junction obtained in the ohmic regime remains of the order of a few GΩ, much larger than the contact resistance measured previously. Neglecting the voltage drop across the contacts, a voltage of 10 Volts applied over a 10-μm-gap between the electrodes corresponds to an electric field of around 10 kV.cm$^{-1}$, in good agreement with the expected value for which the charge carriers mobility starts saturating in the diamond material due to the charge carriers scattering by optical phonons[34,35].

We now demonstrate that the NV-doped junction between the two graphitic electrodes can be used for magnetometry using the PDMR signal (Figure 3). We apply a static magnetic field by approaching a NdFeB permanent magnet mounted on a three-axis translation stage. This magnetic field then removes the degeneracy between the $m_s = \pm 1$ sublevels due to the Zeeman effect.[36] We use this structure to apply to the junction a controlled magnetic field along different crystallographic directions. The associated spin transitions between the NV$^-$ spin states $m_s = 0$ and $m_s = \pm 1$ are driven by a MW magnetic field created by a 1 mm-diameter loop antenna positioned at the top of the junction which is connected to a MW signal generator (SMB100A, R&S) (Figure 3a). The junction is polarized by a 60-V voltage bias corresponding to the saturation regime of the current-voltage characteristics (Figure 2d), then limiting the change in the photoconductivity that could be induced by an electric noise. We first investigate the CW photoconductive response in the absence of magnetic field. As shown in Figure 3b, we observe a decrease of the photocurrent centered at a MW frequency of 2.87 GHz corresponding to the zero-field splitting between the $m_s = 0$ and $m_s = \pm 1$ states of the NV$^-$ charge state[25]. The negative and global contrast of the signal is ~ 12%, close to the value reported by Murooka *et al.* for CW-PDMR experiments implemented in a P-I-N structure[8].

We now apply a static magnetic field along a <100> direction, corresponding to a tilt angle of 54.5° identical for all NV centers in the diamond sample (Figure 3b). In this configuration, the response is identical for the four different families of NV centers and the resonance spectrum consists of two peaks, corresponding to the transitions from the $m_s = 0$ to $m_s = -1$ and from $m_s = 0$ to the $m_s = +1$ spin states. Different spectra are obtained using a 100-mW optical power and a 25-dBm microwave power. From the peak positions, we extract the magnetic field strength applied to the NV-doped diamond junction assuming a <100> direction of the magnetic field[36]. The resonance linewidth of about 20 MHz results mainly from the optical power broadening that occurs in the CW regime[37], since the optically induced electron-spin decoherence rate is generally larger than the Rabi frequency of the MW driving field[3]. The slight decrease in the photoconductive contrast observed when the magnetic field increases is attributed to the influence of the increasing transverse component of the magnetic field[23]. We then study how, for zero magnetic field, optical pumping affects both the photocurrent collected by the graphitic electrodes and the PDMR contrast (Figure 3c). Due to the two-photon ionization process, its intensity (red curve) evolves quadratically with the optical power below the saturation of the excited state $^3E$ to ground state $^3A_2$ radiative transition, and linearly beyond this value. According to the model developed by Siyushev *et al.*[9,28], the photocurrent $I_{NV}$ reads as:

$$I_{NV} = \frac{\frac{P^2}{\alpha\beta}}{1 + \frac{P}{\alpha}} \tag{3}$$

where $\alpha$ (W) and $\beta$ (W.A$^{-1}$) are constants and $P$ (W) is the optical power used to excite the NV centers in the junction. As illustrated on Figure 3c, our measurements are in good agreement with this model. The PDMR contrast (blue curve) plotted in Figure 3c represents the average individual contrast obtained for one of the four NV center families under the qualitative assumption that each NV center family contributes equally to the contrast. We observe that the PDMR contrast decreases as the optical pumping rate increases. We attribute this effect to the competition that occurs in CW regime between the optically induced electron-spin polarization rate and the MW induced Rabi oscillation between the spin states[37] as well as to the increasing of the background photocurrent induced by other impurities such as non-converted nitrogen[3] atoms (e.g. $N_S^0$ defect).

Finally, we applied the magnetic field along a <111> direction (Figure 3a and Figure 3d). With this alignment, the magnetic field amplitude can be increased without losing the NV center spin-dependent photophysical properties for the selected NV-family due to the parasitic influence of the transverse magnetic field[23]. This situation corresponds to the same magnetic field tilt of 71° for three families and purely longitudinal for the last one (Figure 3d). It then induces for this family a linear Zeeman shift[36] of the two transition frequencies between $m_s = 0$ to $m_s = -1$ spin states and between $m_s = 0$ to the $m_s = +1$ spin states:

$$f_{m_s=0 \leftrightarrow m_s=\pm 1} = \left| D \pm \gamma_e . B_\parallel \right| \tag{4}$$

where $D = 2.87$ GHz is the zero field splitting, $B_\parallel$ is the magnetic field component parallel to the NV axis, $\gamma_e = 28$ GHz.T$^{-1}$ is the NV center gyromagnetic ratio. We perform PDMR imaging of the NV-doped diamond junction by continuously adjusting the magnet-to-diamond distance up to a 135-mT magnetic field amplitude (Figure 3d). We observe several drops of the PDMR signal associated to the $m_s = 0$ to $m_s = -1$ transition, corresponding to a magnetic field of 51 mT and 102 mT where an anticrossing due a residual magnetic field transverse component occurs both at the NV⁻ excited state (ESLAC, Figure 3d) and ground state (GSLAC, Figure 3d).[38,39] Moreover, we are able to distinguish a faint PDMR signal associated to the $m_s = 0$ to $m_s = -1$ transition in the excited state level.[40,41] These observations confirm the suitability of graphitic electrodes for efficient PDMR implementation.

In conclusion, we showed how FIB irradiation can be exploited to produce a pattern of graphitic electrodes with ohmic behavior on a diamond crystal. We then used these electrodes to monitor the change of conductivity induced by the spin resonances of an NV ensemble, then achieving PDMR without the need of any metallic deposition on the diamond surface using lithographical pattern. This method paves the way to all-carbon quantum diamond sensors with electrical readout, whose fabrication will require only a limited number of steps then improving the sensor reliability.

Acknowledgements

This project has received funding from the European Union's Horizon 2020 research and innovation program under grant agreement No. 820394 (ASTERIQS), the European Union's Horizon Europe research and innovation program under grant agreement No. 101080136 (AMADEUS), the Marie Skłodowska-Curie grant agreement No. 765267 (QuSCo), the QUANTERA grant agreement ANR-18-QUAN-0008 (MICROSENS), the EIC Pathfinder 2021 program under grant agreement No. 101046911 (QuMicro) and the EMPIR project 20IND05 (QADeT). We acknowledge the support of the DGA/AID under grant agreement ANR- 17-ASTR-0020 (ASPEN).

Author contributions

G.V., L.M., M.S., S.M., M. D. F., J.-F.R., and T.D. contributed to the design and implementation of the research, to the analysis of the results and to the writing of the manuscript. M. D. F. and M. S. contributed to the SRIM simulations and the FIB irradiations. M. M. and A. E. synthesized the diamond crystal.

Conflict of Interest

The authors declare no conflict of interest.

Data Availability Statement

The data that support the findings of this study are available from the corresponding author upon reasonable request.

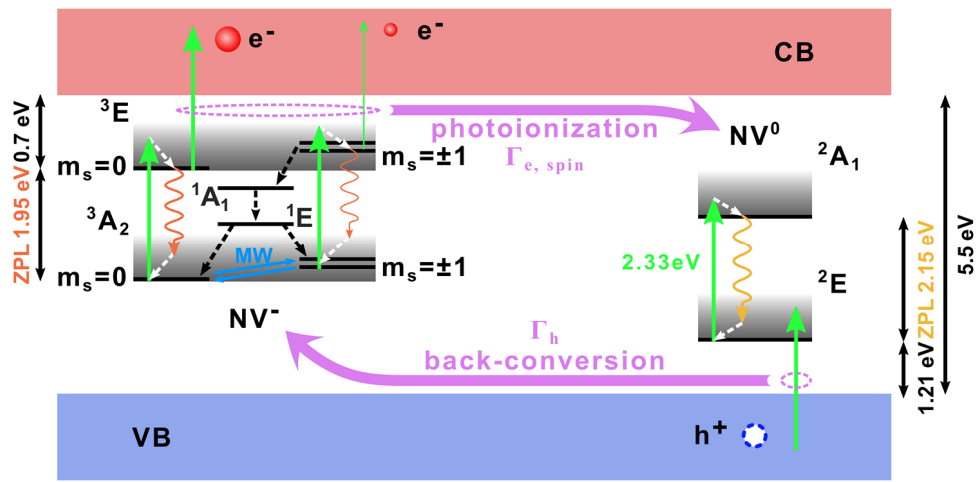

*Figure 1: NV center energy level scheme, for the neutral charge state $NV^0$ and the negatively-charged state $NV^-$, under optical excitation. These discrete levels are inside the bandgap between the valence band (VB) and the conduction band (CB) of the diamond crystal. The solid lines indicate radiative processes whereas the dashed lines indicate non-radiative processes. Due to the spin-dependence of the two-photon induced charge conversion processes between $NV^0$ and $NV^-$, the resonance between the $m_s = 0$ and $m_s = \pm 1$ spin states of the $^3A_2$ level of $NV^-$, induced by a microwave (MW) field, can be equivalently detected by a modification of the photoluminescence intensity or by the photoconductive current associated to the creation of electrons e- with a rate $\Gamma_e$ and holes h+ with a rate $\Gamma_h$.*

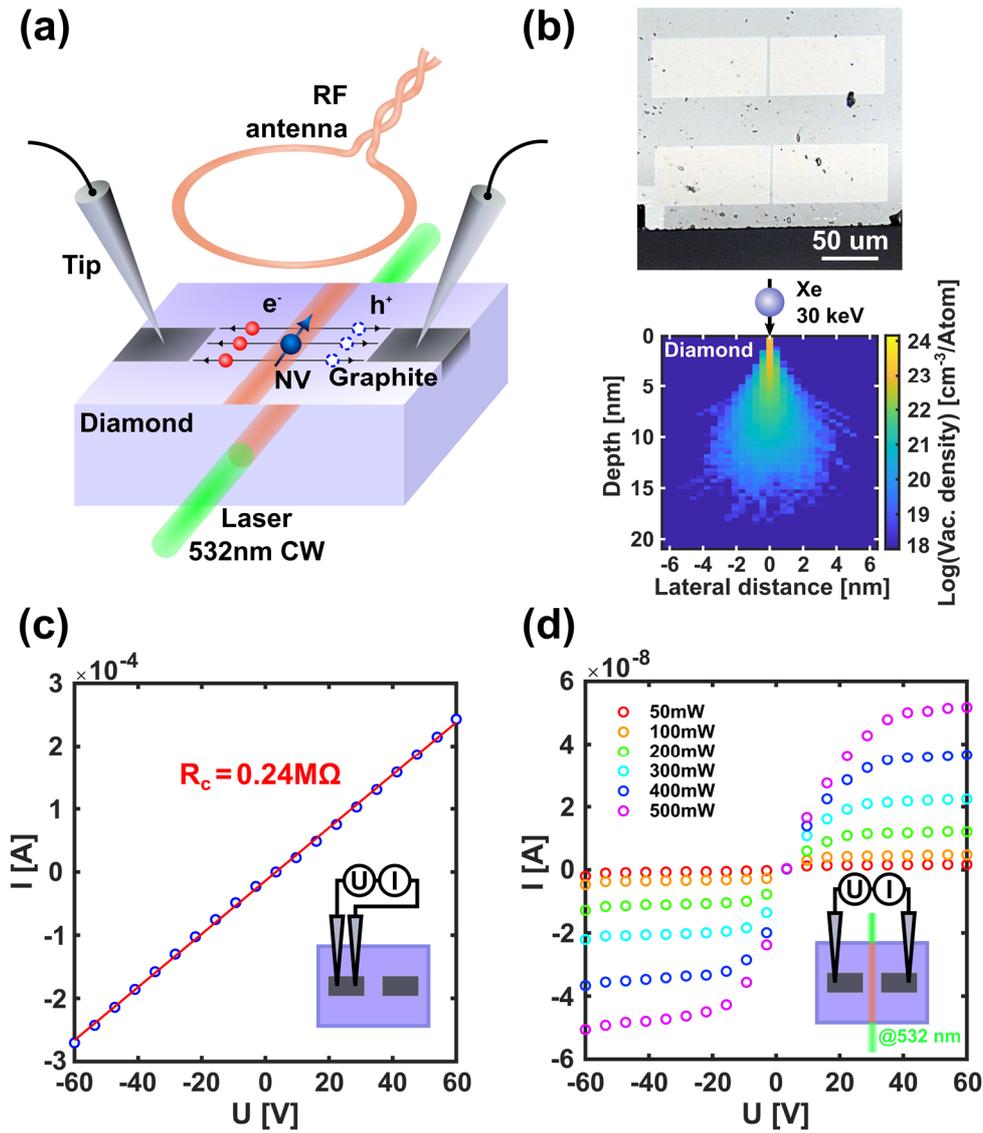

*Figure 2: Photoconductive measurements of the photoinduced charge state conversion $NV^0 \rightleftharpoons NV^-$. (a) Sketch of the experimental setup for the measurement of the current-voltage characteristics of an NV-doped junction between two graphitic electrodes. Two conductive tips are positioned on the electrodes and a green laser is focused in the gap of the junction to induce the spin-selective photoconversion between $NV^0$ and $NV^-$. The spot is about ~ 1.5 µm. (b) Top: Picture of several graphitic electrodes created by FIB irradiation using $Xe^+$ ions. Each electrode has a width of 50 µm, a length of 100 µm and the gaps between two electrodes are 5 µm and 7.5 µm. Down: SRIM simulation of the average profile of the vacancy density with $10^4$ trajectories simulating the penetration of a single $Xe^+$ ion with an incident energy of 30 keV. (c) Current-voltage characteristics measured on the extremities of a graphitic electrode. (d) Same measurements realized across the junction under optical excitation with increasing power.*

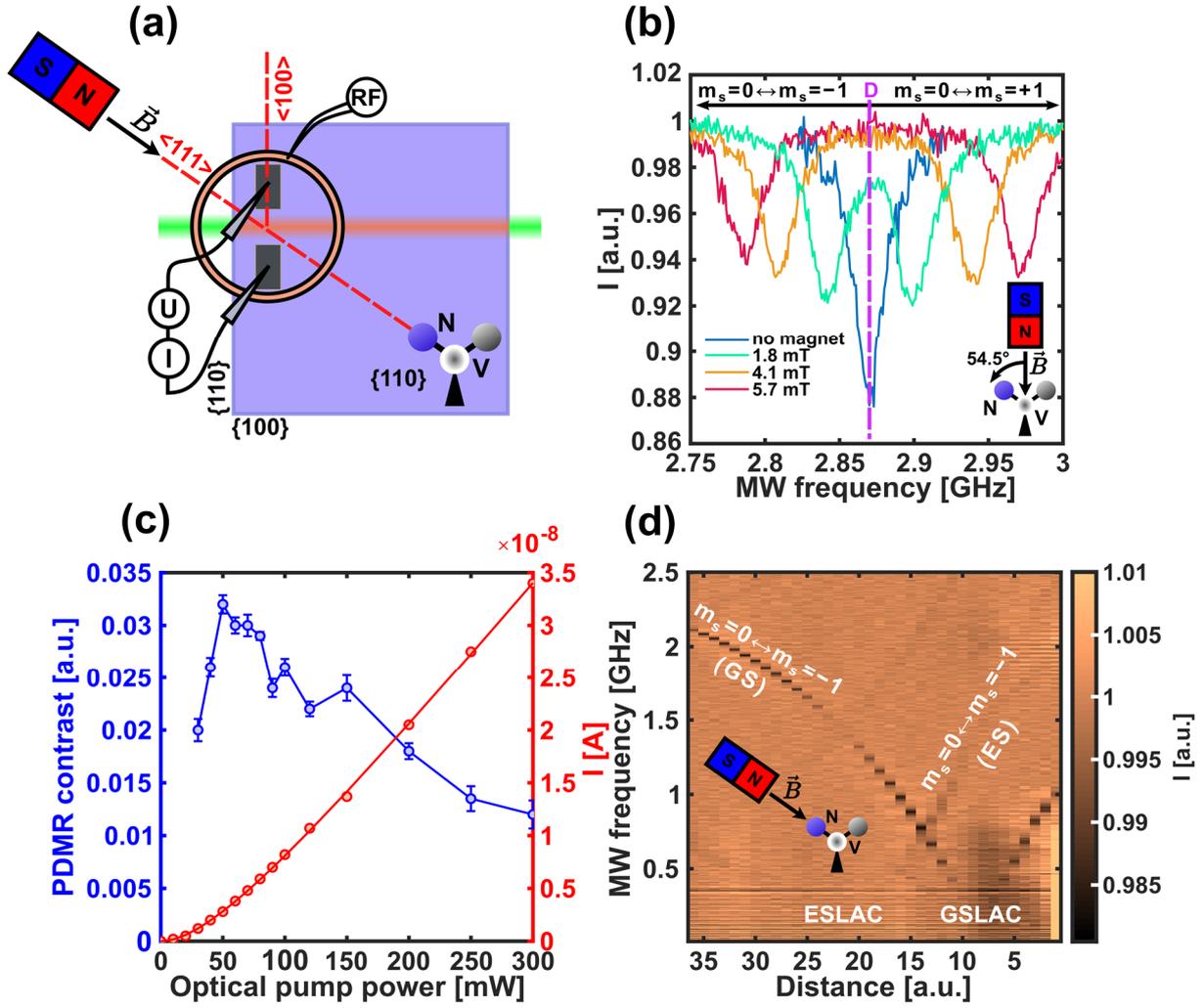

*Figure 3: PDMR implementation on a NV doped diamond junction with graphitic electrodes. (a) Sketch of the experiment indicating the crystal orientation for the face and for both facets. (b) PDMR signal for different magnet to diamond distances. The magnetic field is applied along a <100> axis. (c) Optical power dependence of current collected (red curve) and PDMR contrast (blue) at zero magnetic field. The error bars on the blue curve are due to the error on the fit parameters with a confidence interval of 95% considering a lorentzian model for the PDMR contrast. The solid red curve corresponds to the fit using eq(3) (with: α = 7.04 ± 0.32 W.A$^{-1}$, β = 75,5 ± 6.3 mW at the 95% confidence interval). (d) PDMR signal for different magnet to diamond distances, with a magnetic field applied along a <111> axis. NV$^-$ spin state transition $m_s = 0$ to $m_s = -1$ is observed both at the ground state (GS) and at the excited state (ES). The PDMR contrast vanishes at the anticrossing in both the excited (ESLAC) and ground state (GSLAC) due to a residual transverse magnetic field.*